\documentclass[twocolumn]{aastex631}
\usepackage{amsmath,mathtools,mathrsfs}
\usepackage[caption=false]{subfig}
\usepackage[T1]{fontenc}
\usepackage{hyperref}

\usepackage{color}

\begin{document}
\title{The Image of the M87 Black Hole Reconstructed with \texttt{PRIMO}}

\author{Lia Medeiros}
\affiliation{School of Natural Sciences, Institute for Advanced Study, 1 Einstein Drive, Princeton, NJ 08540}
\affiliation{Steward Observatory and Department of Astronomy, University of Arizona, 933 N. Cherry Ave., Tucson, AZ 85721}
\author{Dimitrios Psaltis}
\affiliation{School of Physics, Georgia Institute of Technology, North Avenue, Atlanta, GA, 30332}
\author{Tod R. Lauer}
\affiliation{NSF's National Optical Infrared Astronomy Research Laboratory, Tucson, AZ 85726}
\author{Feryal Ozel}
\affiliation{School of Physics, Georgia Institute of Technology, North Avenue, Atlanta, GA, 30332}
\correspondingauthor{Lia Medeiros}
\email{lia@ias.edu}

\begin{abstract}
We present a new reconstruction of the Event Horizon Telescope (EHT) image of the M87 black hole from the 2017 data set. We use \texttt{PRIMO}, a novel dictionary-learning based algorithm that uses high-fidelity simulations of accreting black holes as a training set. By learning the correlations between the different regions of the space of interferometric data, this approach allows us to recover high-fidelity images even in the presence of sparse coverage and reach the nominal resolution of the EHT array. The black hole image comprises a thin bright ring with a diameter of $41.5\pm0.6\,\mu$as and a fractional width that is at least a factor of two smaller than previously reported. This improvement has important implications for measuring the mass of the central black hole in M87 based on the EHT images.

\end{abstract}

\keywords{accretion, accretion disks --- black hole physics --- Galaxy: center --- techniques: image processing}
 
\section{Introduction}\label{sec:intro}

The Event Horizon Telescope (EHT) 2017 observations provided high-sensitivity data over long baselines and resulted in the first horizon-scale images of the black hole in M87  \citep{EHTI2019,EHTII2019,EHTIII2019,EHTIV2019,EHTV2019,EHTVI2019} and of Sagittarius A$^*$, the Galactic Center black hole \citep{EHTI2022,EHTII2022,EHTIII2022,EHTIV2022,EHTV2022,EHTVI2022}.
The exceptional resolution achieved by the EHT is made possible by an array of telescopes spanning the Earth and operating as a very long baseline interferometer (VLBI, \citealt{EHTII2019, EHTIII2019}). Despite this global reach, the sparse interferometric coverage of the EHT array (especially during the 2017 observations that have been used for all of the publications to date) makes the already complex problem of interferometric image reconstruction particularly challenging. In such situations, special care is needed to assess the impact of imaging algorithms and sparse interferometric data on the final set of images that can be reconstructed from it. 

A cornerstone of the EHT data analysis strategy was the use of several independent analysis methods, each with different priorities, assumptions, and choices, to ensure that the EHT results were robust to these differences. The use of several general purpose imaging algorithms, for example, was motivated by a desire to reconstruct an image that was consistent with the EHT data while remaining model-agnostic. Those algorithms did not assume ring-like images and could easily have reconstructed a broad range of morphologies. Extensive care was taken by the EHT collaboration to rigorously demonstrate that the ring morphology was uniquely required by the data \citep{EHTIV2019}.

The robustness of the ring-like shapes of the images generated with model-agnostic methods motivates the use of principal-component interferometric modeling (\texttt{PRIMO}), a novel image-reconstruction algorithm that addresses the challenges of millimeter-wave interferometry with sparse arrays by training the algorithm on an extensive suite of simulated images of accreting black holes \citep{Medeiros2022}. In this approach, we apply principal components analysis (PCA) to a large library of high-fidelity, high-resolution general relativistic magnetohydrodynamic (GRMHD) simulations and obtain an orthogonal basis of image components. \texttt{PRIMO} then uses a Markov Chain Monte Carlo (MCMC) approach to sample the space of linear combinations of the Fourier transforms of a number of PCA components while minimizing a loss function that compares the resulting interferometric maps to the EHT data (see \citealt{Medeiros2022} for details on \texttt{PRIMO} and \citealt{Medeiros2018} for an earlier exploration of PCA applied to GRMHD simulations).

General purpose imaging algorithms, such as those used in prior EHT publications, rely on regularizers that, e.g., maximize entropy, minimize gradients, require positivity, and/or prefer compact sources, in order to fill the regions of the Fourier domain where there are no data. In contrast, PCA finds correlations between different regions in Fourier space in the training data, which allows \texttt{PRIMO} to generate physically motivated inferences for the unobserved Fourier components\footnote{Note that, although we perform PCA in the image domain and then take the Fourier transform of the components, the result would have been identical had we performed PCA on the Fourier transform of the images directly~\citep{Medeiros2018}.}. The PCA components themselves contain both positive and negative values. Even though we do not impose positivity on the final image, it emerges naturally in the \texttt{PRIMO} reconstructions.

In machine learning terms, the use of PCA to characterize the GRMHD training set as a sparse orthogonal basis is an example of dictionary learning (see, e.g., \citealt{Shao2014} for a review of dictionary learning applied to image de-noising). Although, in general, decompositions used for dictionary learning do not need to be orthogonal or sparse, PCA does in fact lead to such a decomposition and enables an efficient dimensionality reduction, i.e., requiring a small number of components to fit the data. 

In this Letter, we employ \texttt{PRIMO} to reconstruct the image of the black hole in the center of the M87 galaxy based on the 2017 EHT data. As our training set, we use the simulation library and PCA basis employed in \citet{Medeiros2022} that was based on over 30,000 high-resolution images with a broad range of image morphologies (the resolution of the simulated images is $\lesssim 0.5\mu\rm{as}$, high enough to avoid deleterious aliasing effects; see \citealt{2020arXiv200406210P}). We present the set of images we obtain with this approach and compare them to those reconstructed with agnostic general-purpose imaging algorithms. 

\section{\texttt{PRIMO} Analysis of 2017 M87 data}\label{sec:results}

The data we use in this analysis consists of the EHT observations of M87 taken on April 5, 6, 10, and 11 of 2017. These observations included seven stations at five geographic locations: the Atacama Large Millimeter/submillimeter Array (ALMA), Atacama Pathfinder Experiment telescope (APEX), the James Clerk Maxwell Telescope (JCMT), the Submillimeter Array (SMA), the Arizona Radio Observatory Sub-Millimeter Telescope (SMT), the IRAM 30 m (PV), and the Large Millimeter Telescope Alfonso Serrano (LMT). We choose the April 11th observations as our fiducial data set because it contains a high number of scans as well as good baseline coverage (note that April 5, 6, 10, and 11 contained 18, 25, 7, and 22 scans, respectively). 

Although the data set used for the analysis was first published in \citet{EHTI2019,EHTII2019,EHTIII2019,EHTIV2019,EHTV2019,EHTVI2019}, the EHT collaboration performed additional calibration of the 2017 data prior to the publications on the Galactic Center black hole (Sagittarius A*, Sgr A*, \citealt{EHTI2022,EHTII2022,EHTIII2022,EHTIV2022,EHTV2022,EHTVI2022}). We use this most recent version of the 2017 data, scan averaged and \textit{a priori} calibrated, throughout the manuscript. Because of the \textit{a priori} calibration, we use gain priors that are peaked at unity with a width of only $10\%$. We fix the zero-baseline flux to be 0.6 Jy for consistency with previous EHT analyses, which argued that only a fraction of the observed zero-baseline flux could be attributed to the compact source and the rest was due to extended emission that is not part of the reconstructed image (see the discussion in Appendix B of \citealt{EHTIV2019}).

\begin{figure*}[t!]
\centering
\includegraphics[width=\textwidth]{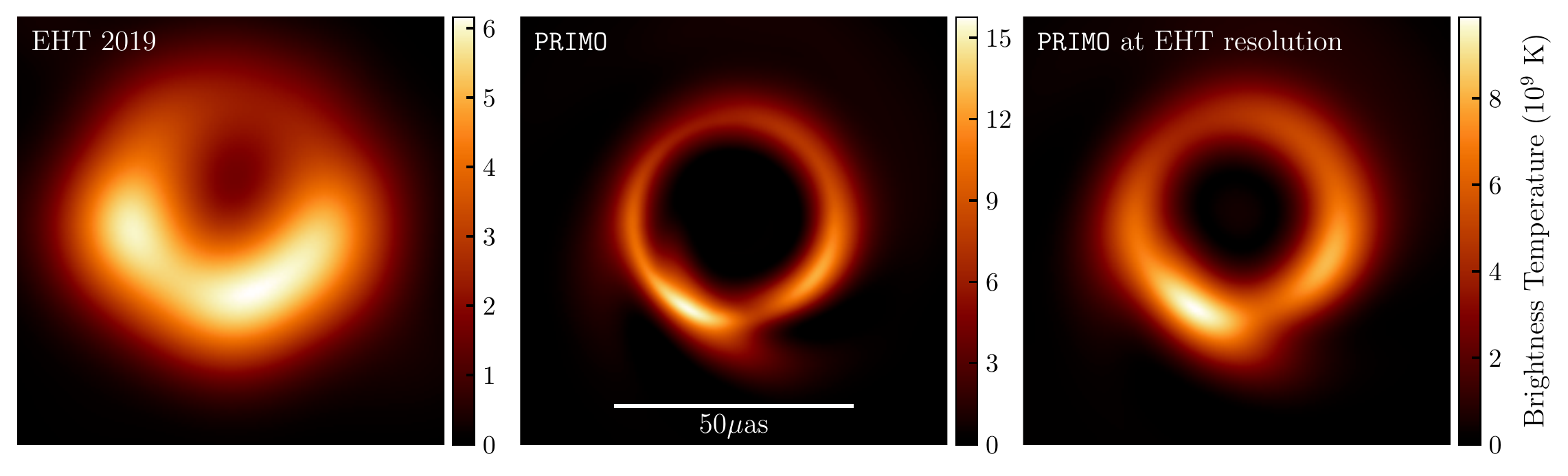}
\caption{\textit{(left)} EHT image of the black hole in the center of M87 based on 2017 data, as reported in \cite{EHTI2019}. \textit{(middle)} Result of reconstructing the image by applying \texttt{PRIMO} to the same data set. \textit{(right)} The \texttt{PRIMO} image blurred to the resolution of the EHT array. The diameter of the ring of emission, the North-South brightness asymmetry, and the central brightness depression are present in all images. The \texttt{PRIMO} image offers a superior utilization of the resolution and dynamical range of the EHT array.}
\label{fig:image}
\end{figure*}

\texttt{PRIMO} optimizes for the ratio of the amplitudes of PCA components to the amplitude of the first component (a total of $N-1$ parameters where $N$ is the number of PCA components used in the reconstruction), an overall size scaling of all components, and an overall rotation of the components on the plane of the sky ($\phi$). The size scaling of the image is expressed in terms of the dimensionless mass-to-distance ratio $\theta_g \equiv GM/Dc^2$, where $G$ is Newton's gravitational constant, $c$ is the speed of light, $M$ is the mass of the black hole, and $D$ is the distance to the black hole.

All images in the training set have the same black hole spin axis, which we assumed to be aligned with the large-scale jet that is observed at longer radio frequencies, pointing toward us at $\sim 17^\circ$ away from the line of sight (see \citealt{2018ApJ...855..128W}). However, we include the possibility of the spin axis pointing away from us at $17^\circ$ by allowing for an overall mirroring of the PCA components, i.e., both clockwise and counter-clockwise accretion flow rotations are included in our model. In \citet{Medeiros2022}, we show that this basis can accurately reconstruct simulated images of black holes with a spin magnitude that is different from the training set, illustrating both the versatility of the training set as well as the small effect of spin on image morphology.

We blur our training set of simulated images before performing PCA so that the decomposition is not overwhelmed by small scale structures that the EHT is not sensitive to. In other words, blurring ensures efficient dimensionality reduction. We use a Butterworth filter with $n=2$ and a scale $r=15\;$G$\lambda$, such that the resulting suppression in the visibility amplitude at baseline lengths the EHT observes is only $\sim 1\%$ and, therefore, smaller than the systematic errors in the measurements (see \citealt{butterworth1930}, \citealt{2020arXiv200406210P}, \citealt{Medeiros2022} for details). This scale of the filter leaves the training set with some power at baselines longer than those of the observations.

The level of resolution in the images obtained with \texttt{PRIMO} depends partially on the number of PCA components used for image reconstruction. We choose 20 based on the optimization performed in \citet{Medeiros2022}, who used several different synthetic data sets and showed that reconstructions with 20 components yield results that are both accurate (minimal biases) and reliable (minimal spurious features), even when the synthetic data originates from simulations with parameters that are distinct from the training set.

The center panel of Figure~\ref{fig:image} shows the image reconstructed with a linear combination of 20 PCA components with \texttt{PRIMO} from 11 April 2017 EHT data of M87. The left panel in the same figure shows the image published in 2019. The salient image features, i.e., the presence of a bright ring of emission and a central brightness depression, the ring size, and the North-South brightness asymmetry are consistent between the two images. The image in the left panel was blurred with a $20\,\mu\mathrm{as}$ Gaussian kernel to mimic the finite resolution of the array. In order to represent the effects of the nominal EHT resolution on the PRIMO image, we apply a $n=2,\, r=8\,\mathrm{G}\lambda$ Butterworth filter to the fiducial image and show the result in the right panel. Contrary to the Gaussian kernel, which suppresses power at intermediate baselines with high signal-to-noise EHT data, the Butterworth filter suppresses power only at baselines where there are no data and approximates more faithfully the maximum resolution of the EHT 2017 array \citep{2020arXiv200406210P}.

The striking difference between the left and right panels of Figure~\ref{fig:image} is the substantially narrower width of the ring and the ability to discern finer features in the \texttt{PRIMO} image. The improved resolution is due to the combination of our use of machine learning to fill in Fourier space as well as our use of a Butterworth filter instead of a Gaussian filter to blur the image.
\texttt{PRIMO} can recover accurate images down to the nominal resolution of the array, where the interferometric data are particularly sparse, because it learns the correlations between the low-frequency and high-frequency structure from the simulations.

\begin{figure}[t!]
\centering
\includegraphics[width=.9\columnwidth]{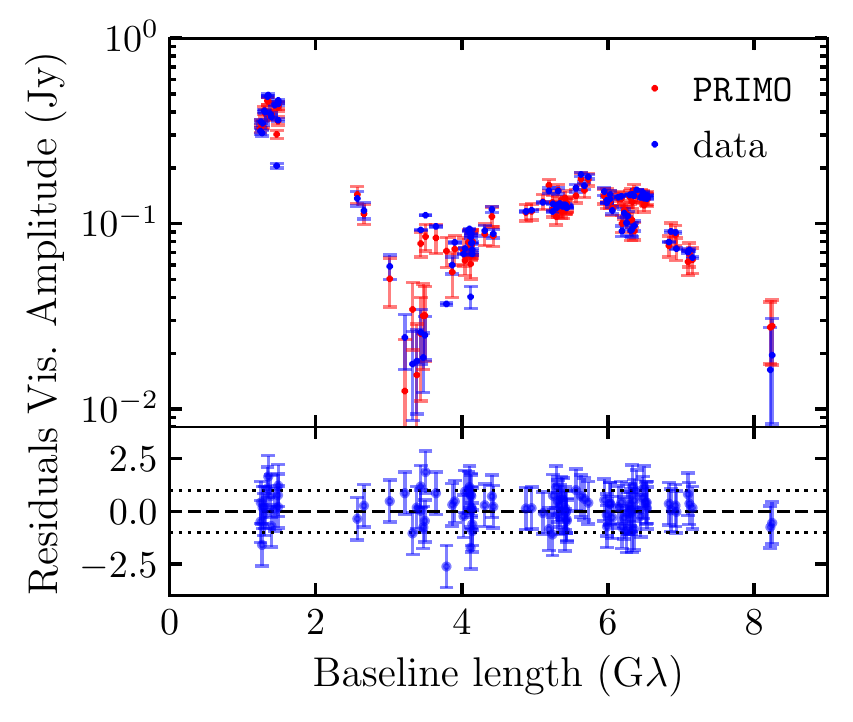}

\includegraphics[width=.9\columnwidth]{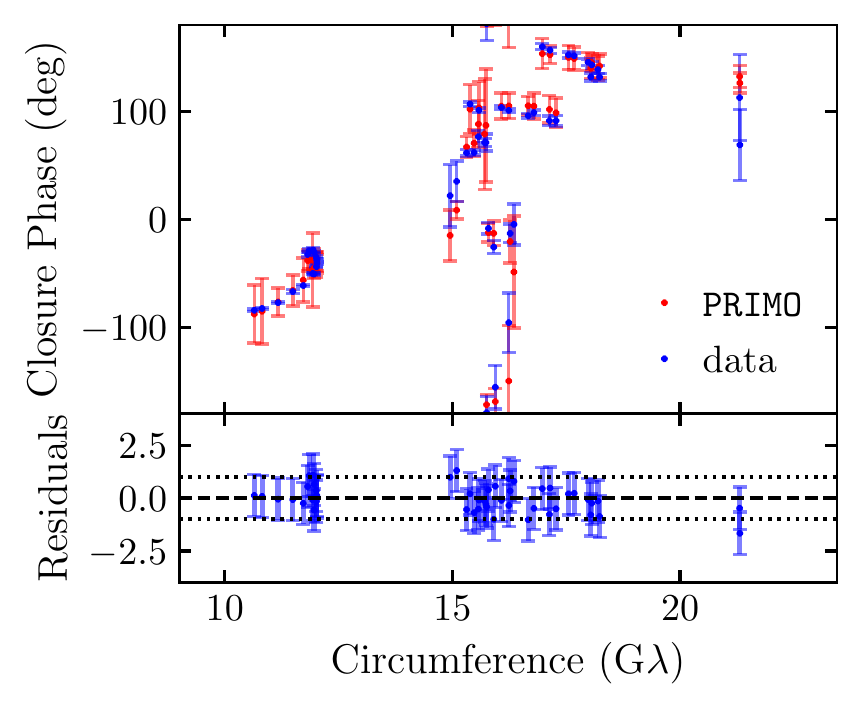}
\caption{Comparison of the visibility-amplitude \textit{(top)} and closure-phase \textit{(bottom)} data to the highest-posterior image reconstruction. The theoretical errors are shown as error bars on the \texttt{PRIMO} reconstruction. Very little structure is seen in the residuals of both the amplitude and closure phase plots.}
\label{fig:amp}
\end{figure}

In Figure~\ref{fig:amp} we compare the EHT data to Fourier amplitudes and closure phases of the highest-posterior image obtained with \texttt{PRIMO}. The match of the \texttt{PRIMO} reconstruction to the observations is excellent, with little structure present in the residuals. Negative values in the highest-posterior image are at the noise level and the total negative flux is only about $1\%$ of the total flux. The theoretical error bars represent the uncertainty introduced by truncating the PCA basis to only 20 components and have been reported in (\citealt{Medeiros2022}, see Fig.~8 and Section 3.5). 

\begin{figure}[t!]
\centering
\includegraphics[width=.9\columnwidth]{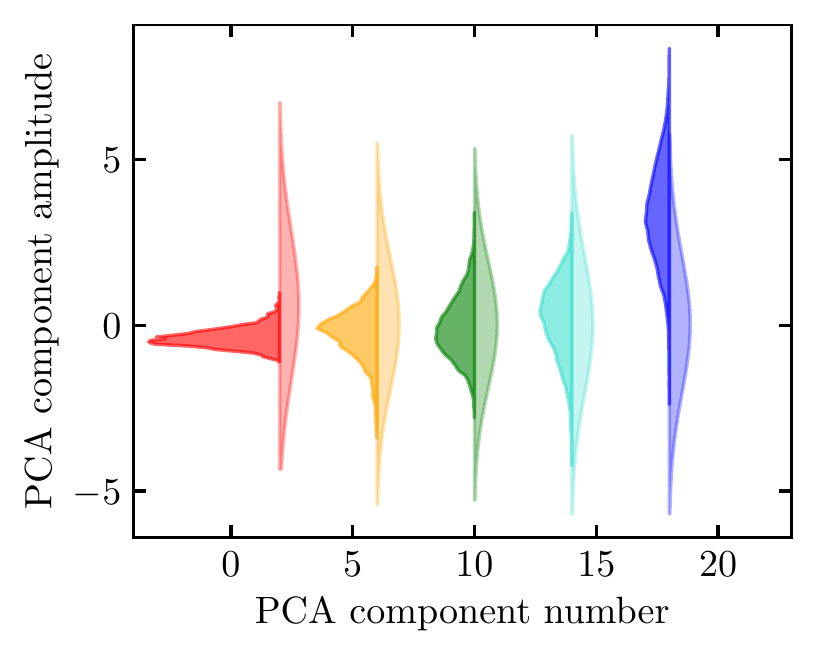}
\caption{Violin plot comparing the posterior \textit{(left half of the violins)} of the amplitudes of PCA components 2, 6, 10, 14, and 18 to their priors \textit{(right half of the violins)}. The posterior distributions of the low-order components are thinner than the priors, indicating that their PCA amplitudes are well constrained by the data.}
\label{fig:violin}
\end{figure}
 
 Figure~\ref{fig:violin} compares the MCMC posterior distribution of the amplitudes of several PCA components to their respective priors. The posterior distributions of the PCA amplitudes of the low-order components are constrained more tightly than the priors, indicating that the amplitudes of the PCA components are well constrained by the data. For the higher-order components, the widths of the posteriors become closer to those of the priors, demonstrating the diminishing power of the data to constrain those low-variance components and justifying our truncation of the PCA basis.
 
\begin{figure*}[t!]
\centering
\includegraphics[width=.9\textwidth]{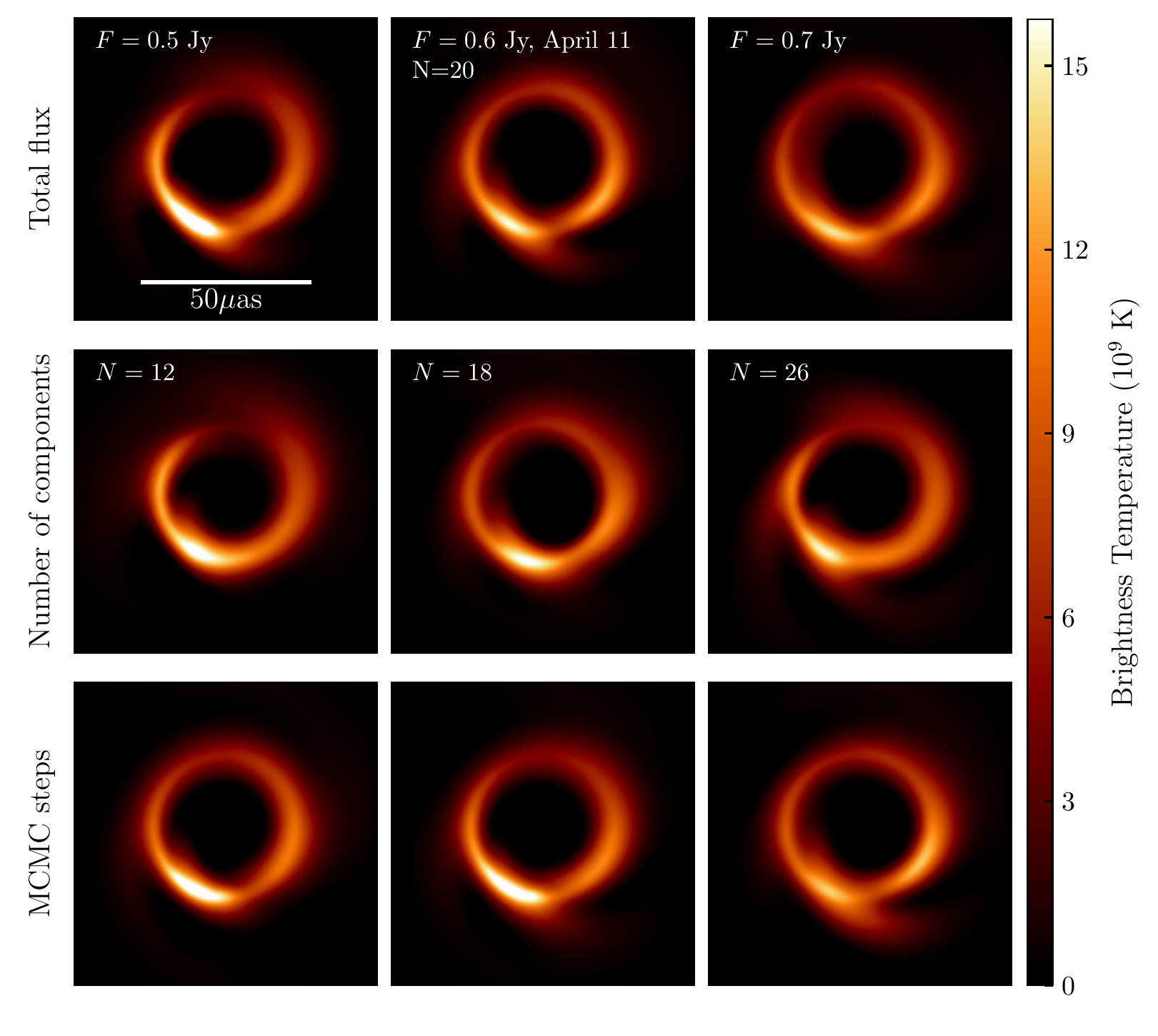}
\caption{\textit{(top)} Comparisons of highest-posterior \texttt{PRIMO} images with total flux equal to 0.5, 0.6, and 0.7 Jy. \textit{(middle)} Comparisons of the highest-posterior images using only 12, 14, and 18 PCA components. The morphology of the image is robust to changes in total flux and the number of PCA components. \textit{(bottom)} Example images randomly drawn from the MCMC steps of the fiducial chains with flux of 0.6 Jy and 20 PCA components. All three example images have posteriors that are in the 50th percentile.}
\label{fig:survey}
\end{figure*}

\section{Parameter Study}\label{parameter}

The fiducial \texttt{PRIMO} image of M87 using the 2017 EHT data discussed in the previous section sets the total flux of the compact source to 0.6 Jy, and reconstructs the image with a linear combination of 20 PCA components. In Figure~\ref{fig:survey} we compare our fiducial image to those obtained with different total compact source flux and different numbers of PCA components. The main image features, i.e. the presence and size of the ring, the presence and depth of the brightness depression, and the orientation angle of the brightness asymmetry along the ring are robust to changes in both the total compact flux and the number of PCA components used. All fits prefer a black hole spin axis pointing away from the observer at $17$ deg away from the line of sight, consistent with the results of \citet{EHTV2019}. 

The extended feature towards the bottom of the fiducial image is present in several of the fits in Figure~\ref{fig:survey}, but is not robust to changes in the number of PCA components. The bottom row of Figure~\ref{fig:survey} shows three example images randomly drawn from the top 50th percentile of the MCMC chains of the fiducial reconstruction. The relative brightness of this feature is different between these images, indicating a significant uncertainty. Although this feature may be real, and would be consistent with features commonly seen in simulations, it is not required by the data. The relative brightness of the brightest part of the ring and its position angle are also weakly constrained. This uncertainty arises from the sparse baseline coverage of the 2017 EHT array, which makes constraining the precise azimuthal structure challenging (see, e.g. Figure 3 of \citealt{2022ApJ...928...55P} and the accompanying discussion). Such differences are also in images obtained with other algorithms with the same data set (see e.g., Figures 4, 6, 7, and 8 of \citealt{EHTIV2019}). Critically, however, the ring size and width are robust in all \texttt{PRIMO} images, and the ring is always brighter towards the South, owing to the tight constraints from the closure phases measured in the smallest baseline triangles.

\begin{figure*}[t!]
\centering
\includegraphics[width=\textwidth]{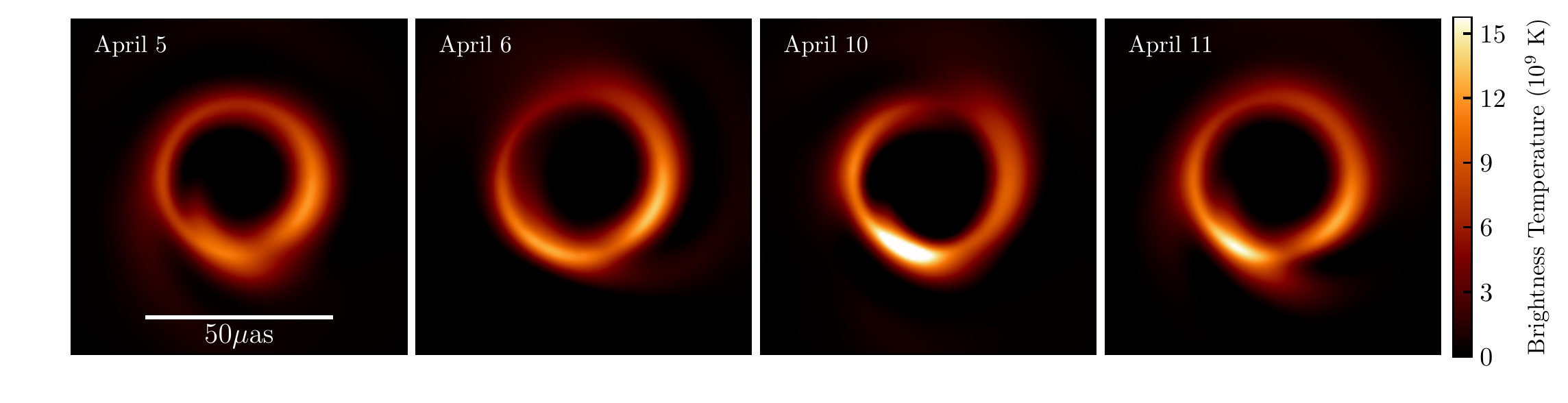}
\caption{Comparisons of reconstructed images based on the EHT data obtained on 2017 April 5, 6, 10, and 11. All reconstructions were performed with 20 PCA components and a compact flux of 0.6~Jy. The diameters and widths of the images are consistent across the duration of the 2017 observing campaign, which spans  approximately one dynamical timescale for phenomena near the horizon of the M87 black hole.}
\label{fig:days}
\end{figure*}

We also reconstructed images with 20 PCA components from the EHT observations taken on April 5th, 6th, and 10th, setting the total compact flux to 0.6 Jy. The baseline coverage of April 10th is significantly poorer than that of the other days, with April 6th and 11th having the best coverage (see, e.g., Figure 1 in \citealt{EHTIV2019}). \texttt{PRIMO} can reconstruct high-resolution images of all days (see Figure~\ref{fig:days}), with relatively minor changes between days. The orientation angle of the brightest part of the ring, the relative ring brightness asymmetry, and the brightness and location of the southern extended feature are slightly variable between days. Some of these changes can be attributed to the difference in baseline coverage and errors, as discussed earlier. However, some real differences between closure phase data of the first two days and the last two were identified in \citet[see e.g. Figure 3]{EHTIV2019}. Therefore, we conclude that the differences between the first two and the last two images, such as the changes in orientation angle of the brightest part of the ring and the ring brightness asymmetry, are likely caused by observed differences in the source structure. 

\section{Discussion}\label{conclusions}

We present a new image reconstruction of the M87 supermassive black hole based on \texttt{PRIMO}, which uses dictionary learning to correct for the sparse Fourier-domain coverage of the EHT interferometric visibilities. This approach relies on using a large library of synthetic images from general relativistic magnetohydrodynamic simulations of accreting supermassive black holes as a training set. It optimizes the benefits of an agnostic image reconstruction technique by remaining flexible enough to reproduce a broad range of morphologies while using physically motivated training (as opposed to ad hoc regularizers) to infer the maximum information from the data. 

The image of the M87 black hole that we have reconstructed with \texttt{PRIMO} is dominated by a narrow ring that has a substantial brightness depression at its center and a rim that is brighter towards the South. The ring-like shape is a consequence of the observed dependence of the interferometric visibility amplitudes on baseline length. As the top panel of Figure~\ref{fig:amp} shows, this dependence has the characteristic shape of a Bessel function, which is the Fourier transform of a ring-like image. For an infinitesimally thin ring, the baseline length $b_1\simeq 3.75$~G$\lambda$ that corresponds to the first visibility amplitude minimum is directly related to the diameter of the ring as
\begin{equation}
d_0\simeq 42 \left(\frac{b_1}{3.75~{\rm G}\lambda}\right)^{-1}~\mu{\rm as}\;.
\label{eq:diameter}
\end{equation}
In order to measure directly the diameter of the ring-like images reconstructed here with \texttt{PRIMO}, we randomly sampled 1,750 steps from the \texttt{PRIMO} MCMC chains of the fiducial image and applied the \texttt{CHARM} feature extraction algorithm~\citep{Ozel2022,EHTVI2022}. We find the ring diameter to be $41.5\pm 0.6\,\mu$as, which is in good agreement with the estimate based on equation~(\ref{eq:diameter}). Naturally, the measured image diameter we report here also falls within the range of 39-45~$\mu$as inferred using other imaging and visibility-domain model fitting algorithms across the various observing days~\cite{EHTI2019,EHTVI2019}.

The width of the ring-like image is another important diagnostic that can be used to constrain physical characteristics of the accretion flow such as the accretion rate. Moreover, the fractional ring width provides a natural upper bound on the potential difference between the measured diameter of the image and the diameter of the black-hole shadow and, hence, on the uncertainty in the inference of the black-hole mass. Indeed, extensive simulations have shown that the outline of the black-hole shadow is contained within the width of the ring-like image, even though it might not coincide with the location of maximum brightness, independent of the plasma properties in the radiatively inefficient flow or the underlying metric of the spacetime~\citep{Ozel2022,Younsi2021}. 

In practice, the width of a ring-like image is determined by the visibility amplitudes at and beyond the baseline lengths that correspond to the first bump of the top panel of Figure~\ref{fig:amp}, i.e., beyond $\sim 5$~G$\lambda$. The sparseness of the interferometric data at these large baseline lengths hampers substantially the ability of traditional imaging algorithms to constrain ring widths. In the case of the regularized maximum likelihood algorithms, this sparseness of data causes the inferred ring width to be determined primarily by the strengths of the regularizers, which are necessary to bound the otherwise underdetermined image reconstruction (see, e.g., Fig.~7 of \citealt{EHTIV2019}). As such, only a conservative upper bound of $\lesssim 0.5$ fractional width has been reported so far based on earlier imaging algorithms~\citep{EHTI2019}.

In contrast, training \texttt{PRIMO} with the large suite of synthetic images obviates the need for regularizers and results in reconstructions that reach the nominal resolution of the EHT array. 
We find that the ring width in the same sub-sample of MCMC chains of \texttt{PRIMO} images is $9.6\pm0.5\,\mu$as, which corresponds to a fractional width of $\lesssim 0.23$ that is a substantial improvement compared to the earlier inference. This improvement will lead to reduced errors in the inferred mass of the M87 black hole based on the reconstructed image. Achieving this, however, requires a careful calibration between the diameter of the ring measured in the images and that of the black hole shadow boundary.  We leave a quantitative calibration of our algorithm (similar to that performed in \citealt{EHTVI2022}) to future work.

Future EHT observations with additional telescope locations and higher bandwidth will also allow us to use a higher number of PCA components with \texttt{PRIMO} with higher variance in the azimuthal direction. This will lead to further improvements in the effective resolution of the image and allow us to better constrain the image morphology. 

\begin{acknowledgements}
The authors thank the anonymous reviewer for comments that improved the manuscript. L.\;M.\ gratefully acknowledges support from an NSF Astronomy and Astrophysics Postdoctoral Fellowship, award no. AST-1903847. D.\;P.\, and F.\;O.\, gratefully acknowledge support from NSF PIRE grant 1743747.
\end{acknowledgements}


\bibliography{main}
\end{document}